\documentclass[twocolumn,showpacs,pra,aps,superscriptaddress,floatfix]{revtex4}
\usepackage{graphicx}
\usepackage{comment}
\usepackage{braket}             		
\usepackage{float}  				    
\usepackage[utf8]{inputenc} 
\usepackage{natbib}
\usepackage{soul}
\usepackage{color}
 \RequirePackage{color}
 \definecolor{darkred}{rgb}{0.8,0.1,0.1}

\newcommand{\beq}{\begin{equation}}
\newcommand{\eeq}{\end{equation}}
\newcommand{\beqa}{\begin{eqnarray}}
\newcommand{\eeqa}{\end{eqnarray}}

\begin{document}

\title{Transient quantum beats, Rabi-oscillations and delay-time of modulated matter-waves}
\author{Jorge Villavicencio}
\affiliation{Facultad de Ciencias, Universidad Aut\'onoma de Baja
California, 22800 Ensenada, B.C., M\'exico}
\email{villavics@uabc.edu.mx}
\author{Alberto Hern\'andez-Maldonado}
\affiliation{Facultad de Ciencias de la Ingenier\'{\i}a y Tecnolog\'{\i}a, Universidad Aut\'onoma de Baja California, Unidad Valle de las Palmas, Tijuana, Baja California, M\'exico}
\email{hernandez.alberto@uabc.edu.mx}
\date{\today}
\begin{abstract}

Transient phenomena of phase modulated cut-off wavepackets are explored by deriving an exact general solution to Schr\"odinger's equation for finite range potentials involving  arbitrary initial quantum states.
We show that the dynamical features of the probability density are governed by a virtual \textit{self-induced two-level system} with energies $E_{+}$, and  $E_{-}$, due to the phase modulation of the initial state.  
The asymptotic probability density exhibits Rabi-oscillations characterized by a frequency $\Omega=(E_{+}-E_{-})/\hbar$, which are independent of the potential profile. 
It is also found that for a system with a bound state,
the interplay between the virtual levels with the latter causes a \textit{quantum beat} effect with a beating frequency, $\Omega$.
We also find a regime characterized by a \textit{time-diffraction} phenomenon 
that allows to measure unambiguously the delay-time, 
which can be described by an exact analytical formula.
It is found that the delay-time agrees with the phase-time only for the case of strictly monochromatic waves.
\end{abstract}
%

\maketitle

\section{Introduction}

Transient phenomena have lead to the development of powerful analytical and experimental techniques to explore the dynamics of matter-waves \cite{kleberPR94, dcgcjm}.
Transient effects emerge as a response of a quantum system to various types of interactions, state preparation or initial boundary conditions.
The most representative example of transient phenomena due to a sudden interaction is the Moshinsky quantum-shutter setup \cite{mm52}.
The latter involves the time-dependent solution of free Schr\"odinger's equation for a cut-off plane wave of momentum $k$, initially confined to the left of a perfect absorbing shutter located at $x=0$.
By opening the shutter at a time $t = 0$, the quantum state evolves freely in space, and the probability density in the time-domain exhibits a distinctive oscillatory pattern known as \textit{diffraction in time} \cite{mm52,mm76}, 
analogous to the intensity profile of a light beam diffracted by a semi-infinite plane.
The time-diffraction effect was verified decades later in experiments using ultracold atoms \cite{dalibard}, cold-neutrons \cite{Hils98}, and atomic Bose-Einstein condensates \cite{colombe05}. 
The quantum-shutter problem has allowed to translate spatial features of light optics to the time-domain of matter-waves in experiments with atoms that exhibit diffraction and interference \cite{Arndt96,dalibard}, as well as phase modulation \cite{steane95,decamps16} and quantum beat phenomena \cite{decamps16}. 

After Moshinsky's proposal, several theoretical works have addressed the study of transients of tunneling matter-waves within a quantum shutter setup for systems involving delta potentials  \cite{kleberPR94,19306687,gcah03,adjpa04,19156601,grmar07,mlgc10}, potential barriers \cite{bm96,ggcjvpra01,12286114,grmar07,Julve_2008}, and multibarrier resonant structures \cite{gcar97,PhysRevB.60.R2142,Cordero_2010}.
Some of these transients involve electronic transport and buildup in quantum structures \cite{gcar97,PhysRevB.60.R2142,Cordero_2010}, and time-scales \cite{kleberPR94,19306687,gcar97,ggcjvpra01,gcah03,adjpa04,19156601,12286114}.  
Although most of these works have relied on cut-off plane wave initial conditions \cite{19306687,gcah03,ggcjvpra01,12286114,adjpa04,bm96,19156601,Julve_2008,mlgc10} and wave packets with Gaussian   distributions \cite{adjpa04, 19156601,Cordero_2010}, to our knowledge, the study and characterization of transient effects of phase modulated cut-off waves has not been addressed before.

In this work we explore  transient phenomena of phase modulated cut-off wavepackets by deriving an exact analytical solution to Schr\"odinger's equation based on a resonant state expansion. 
The modulation involves a superposition of quantum states with slightly different momenta.  
The aim of this work is to show that the phase modulation of initial cut-off quantum waves gives rise to interesting transient phenomena, such as Rabi-oscillations, quantum beats, delay-time, and time-diffraction effects. 
We discuss how this these effects are the result of a virtual \textit{self-induced two-level system} that governs the dynamics of probability density in the time-domain.

Our work is organized as follows. In Sec.~\ref{model} we present the main equations involving the solution of the quantum shutter approach for a general potential to describe the dynamics of arbitrary cut-off initial quantum states. 
In Sec.~\ref{results} we study the time-dependent features of  phase modulated wavepackets along the transmission region of a potential, which  exhibits different regimes governed by a virtual self-induced  two-level system. The effects of the modulation on the delay-time of the system are also explored.  
Finally, in Sec.~\ref{conclusions} we present the conclusions.

\section{Model for a general initial state\label{model}}

Our approach deals with an exact analytical solution to Schr\"odinger's equation for a finite range potential involving an arbitrary cut-off initial state, based on a resonant state expansion. 
We stress that our approach is non-Hermitian since the  resonant states are eigenfunctions of the Hamiltonian with \textit{outgoing boundary conditions}, which leads to complex energy eigenvalues. The Hermiticity of the system Hamiltonian is not only related to the  operator itself but also to the functions on which it acts on \cite{moiseyev_2011}.  Non-Hermitian approaches provide alternative exact analytical descriptions of physical processes. In fact, the equivalence of non-Hermitian dynamical descriptions and those based on \textit{continuum wave expansions} of standard quantum mechanics has been demonstrated in Ref. \onlinecite{12286114}.
Within this framework, the time-dependent features of  quantum waves with different initial conditions can be explored. 
Our model is a generalization of the work originally developed for a cut-off plane wave in Ref.~\onlinecite{gcar97}. 
The general approach to the problem involves the time-evolution of particles of energy $E=\hbar^2 k^2/2m$ incident  from the left onto a one-dimensional finite range potential $V(x)$ of arbitrary shape that extends along the region $0\le x \le L$, and vanishes thereafter.
The  wavefunction $\Psi(x,t)$ of the problem is a solution of 
Schr\"{o}dinger's equation $\hat{H}\Psi(x,t)=i\hbar\partial_t\Psi(x,t)$, with  $\hat{H}=(-\hbar^2/2m)\partial^2_x+V(x)$, for an initial condition at $t=0$ given by, 
\begin{equation} 
\Psi(x,0)=\cases{\phi(x,k), & for $x\leq0;$\cr 0, & for $x> 0$.}
\label{eq1_bis}
\end{equation}
The function  $\phi(x,k)$ represents a general initial state that extends from $-\infty < x \leq 0$. 
The  solution is obtained by Laplace-transforming  Schr\"{o}dinger's equation and the corresponding  initial condition, following the procedure of Ref. \onlinecite{gcar97}, that is,  
\begin{equation}
\left[ \partial^2_x +p^2\right]\widetilde{\Psi}(x,p)=\cases{i\alpha\, \phi(x,k), & for $x\leq 0;$\cr 0, & for $x > L$.}
\label{newdifeq}
\end{equation}
The  solutions to Eq.~(\ref{newdifeq}) are,
\beqa
\widetilde{\Psi}(x,p)=\cases{c^+e^{i p x}, & for $x> L;$\cr c^-e^{-i p x}+ \alpha \widetilde{F}(x,k,p)\, e^{-i p x}, & for $x \leq 0$,}
\label{newdifeq_sol}
\eeqa
where we have defined $p=(i\alpha s)^{1/2}$ with $\alpha=2 m/\hbar$, and  $c^{\pm}$ are constants to be determined by the matching conditions of the wavefunctions.
The function $\widetilde{F}(x,k,p)$ is the particular solution to  the inhomogeneous equation in (\ref{newdifeq}), given by
\beqa
\widetilde{F}(x,k,p)=\int e^{2 i p x}\left[\int \, i\, \phi(x,k) \,e^{-i p x} dx \right] dx,
\label{intinception}
\eeqa
obtained following the procedure of Ref. \onlinecite{JOHNSON2008582}.

\begin{figure}[!tbp]
    \includegraphics[width=3.7in]{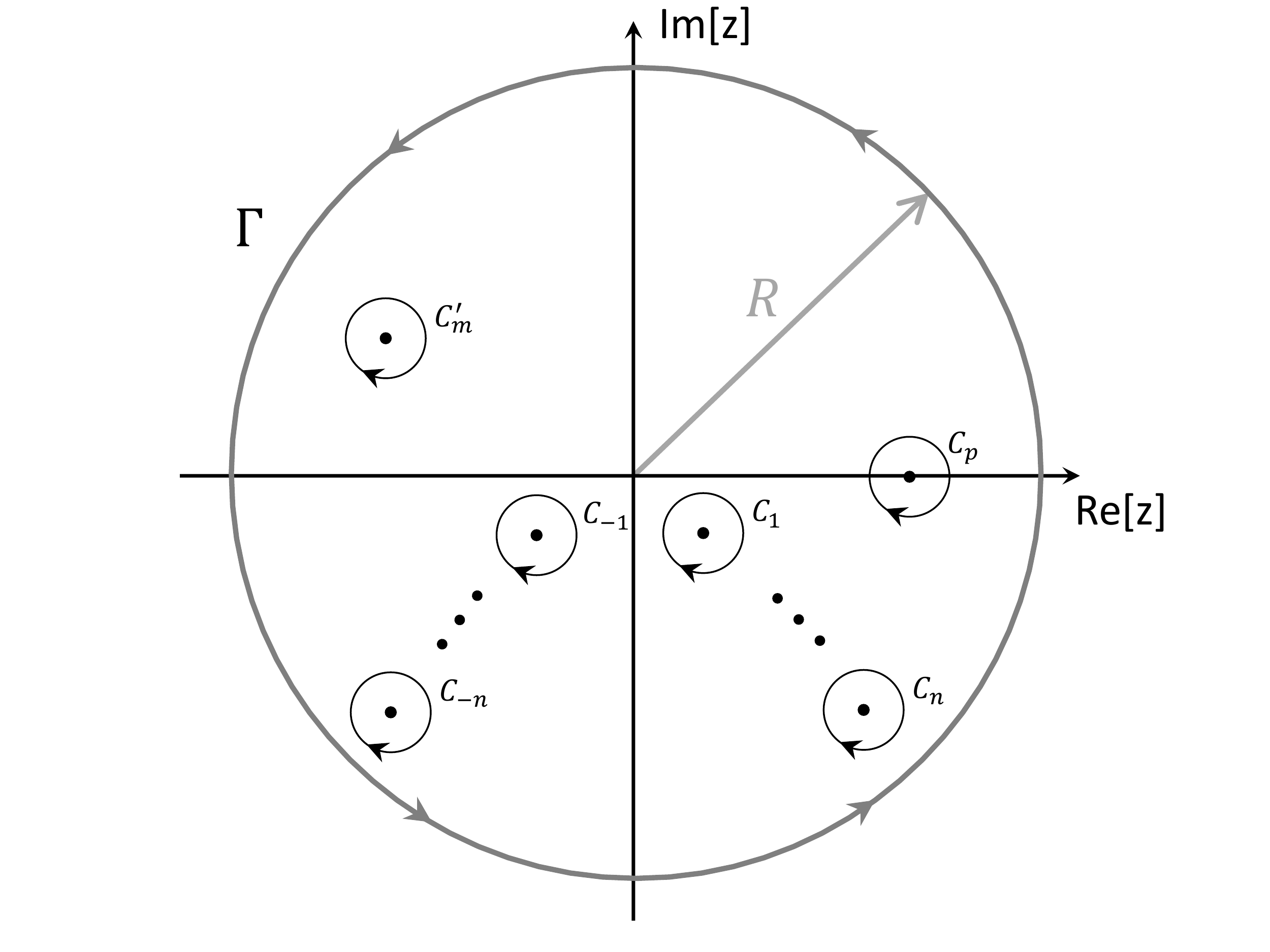}
     \caption{Contour in the complex $z$-plane to solve Eq.~(\ref{cacuchy_1}). The complex poles are enclosed by the following contours: $\Gamma$ of radius $R$ centered about the origin in the $z$-plane following an anticlockwise direction, $C_p$ that encloses a pole at $z=p$, all the $C'_{m}$ that enclose simple poles at $z= \kappa_m$, and all the $C_n$ that encircle the simple complex poles ${k_n}$ of $G^+(x,x';z)$, all in a clockwise direction.} 
      \label{contour}
        \end{figure}

The differential equation for the internal region reads,
\beqa
\left[ \partial^2_x +p^2-\widetilde{V}(x) \right]\widetilde{\Psi}(x,p)=0, \,\,\,\, 0 \leq x \leq L,
\label{newdifeq2}
\eeqa
with $\widetilde{V}(x)=2mV(x)/\hbar^2$. 
We obtain the solution for the internal region by using the corresponding equation of the outgoing Green's function of the system, $G^+(x,x';p)$, that is,
\beqa
\left[ \partial^2_x +p^2-\widetilde{V}(x) \right]G^+(x,x';p)=\delta(x-x'), \,\, 0 \leq x \leq L,
\label{green_new}
\eeqa
with the outgoing boundary conditions,
\begin{eqnarray}
    \partial_x G^+(x,x';p)|_{x=0}&=&-i \,p \,G^+(0,x';p), \label{Gobc1}\\ 
   \partial_x G^+(x,x';p)|_{x=L}&=&i\, p\, G^+(L,x';p).\label{Gobc2}
\end{eqnarray}
We write $\widetilde{\Psi}(x,p)$ in terms of $G^+$ by multiplying Eq.~(\ref{newdifeq2}) by $G^+(x,x';p)$ and Eq.~(\ref{green_new}) by $\widetilde{\Psi}(x,p)$, and subtracting the equations.
Thus, by integrating the resulting expression along  $x\in[0,L]$, with the help of Eq.~(\ref{newdifeq_sol}) and the boundary conditions (\ref{Gobc1}) and (\ref{Gobc2}), we get
\begin{equation}
    \widetilde{\Psi}(x',p)=\alpha\,\widetilde{\Phi}(0,k,p) \,G^+(0,x';p),
\label{psi_interanl1}    
\end{equation}
where $\widetilde{\Phi}(0,k,p)=\partial_x\widetilde{F}(x,k,p)|_{x=0}$.
By evaluating Eq.~(\ref{psi_interanl1}) in $x'=L$, and  Eq.~(\ref{newdifeq_sol}) in $x=L$, we obtain the constant $c^+$ which allows to write the solution along the transmission region ($x \geq L$) as 
\begin{equation}
    \widetilde{\Psi}(x,p)=\alpha\,C(p,k)\,e^{i p x}\,\,\,x \geq L,
\label{psi_interanl1c}    
\end{equation}
with $C(p,k)\equiv \widetilde{\Phi}(0,k,p)\,G^+(0,x';p) \,e^{-i p L}$.
We calculate $C(p,k)$ by evaluating the following complex integral in the $z$-plane using Cauchy's theorem,
\begin{equation}
I=\frac{1}{2\pi i} \int_C \frac{z \,C(z,k)}{z-p}\ dz=0,
\label{cacuchy_1}    
\end{equation}
with $C(z,k)\equiv \widetilde{\Phi}(0,k,z)\,G^+(0,x';z) \,e^{-i z L}$, by considering the integration contour $C=\Gamma+ C_p + \sum_n C_{n}+\sum_m C'_{m}$. See Fig.~\ref{contour}.  The contours $C_n$ encircle the complex poles $k_n$ of $G^+(x,x';z)$ which is a meromorphic function {\it i.e.} the Green function has simple poles. 
The contours $C'_m$ encircle the poles $\kappa_m$ of $\widetilde{\Phi}(0,k,z)$, which is also a meromorphic function. 
The complex integration leads us to, 
\begin{eqnarray}
&&\frac{1}{2 \pi i} \Bigg[ \int_{C_p} \frac{z \,C(z,k)}{z-p}\ dz  \nonumber \\
&+&  \sum_n\int_{C_n} \widetilde{\Phi}(0,k,z)\,G^+(0,x';z) \,e^{-i z L} \ dz  \nonumber \\
 &+&\sum_m\int_{C'_m} \widetilde{\Phi}(0,k,z)\,G^+(0,x';z) \,e^{-i z L}\ dz \Bigg]=0. \nonumber \\
\label{integrales_complejas}
\end{eqnarray}
We perform the integration in Eq.~(\ref{integrales_complejas}) by applying the Cauchy residue formula, and using the following expansions for $\widetilde{\Phi}(0,k,z)$, and $G^+(0,x';z)$. For the case $\widetilde{\Phi}(0,k,z)$ we assume a Mittag-Leffler expansion
\beqa
\widetilde{\Phi}(0,k,z)=\widetilde{\Phi}(0,k,0)+\sum_{m'=1}^{\infty} \alpha_{m'} \left( \frac{1}{z-\kappa_{m'}}+\frac{1}{\kappa_{m'}}\right),
\label{mittaglef}
\eeqa
where $\alpha_{m'}$ are the residues at the simple complex poles $\kappa_{m'}$.
For the case of $G^+(0,x';z)$, we use the expansion of the Green function  in terms of the complex poles $k_n$ and the resonance states \cite{gcaqc10} $u_n(x)$,  namely
\begin{equation}
G^+(x,x';z)=\sum_{n=-\infty}^{\infty}\frac{u_n(x)u_n(x')}{2k_n(z-k_n)},
\label{Ggaston}
\end{equation}
where the $u_n$'s are one-dimensional resonant states (\textit{quasinormal states}), which are
eigensolutions of the Schr\"{o}dinger equation,
\begin{equation}
\frac{d^2u_n(x)}{dx^2}+\left[ k_n^2-\widetilde{V}(x)\right]
\,u_n(x)=0, \label{SEun}
\end{equation}
with outgoing boundary conditions:
\begin{eqnarray}
\frac{du_n(x)}{dx}\Big|_{x=L}&=&+ik_nu_n(L);\nonumber \\
\frac{du_n(x)}{dx}\Big|_{x=0}&=&-ik_nu_n(0). 
\label{boundary1}
\end{eqnarray}
The index $n$ ($n=1,2,3,...$) spans both the third  and fourth quadrant poles of the complex $k$-plane, given respectively by $k_{-n}$ and $k_n$. 
These poles  are related through $k_{-n}=-k_n^*$, which follows from time-reversal invariance.
The complex energies associated to $k_n$ are given by $E_n=\hbar^2k_n^2/2m=\varepsilon_n-i\Gamma_n/2$, where $\varepsilon_n$ is the resonance energy, and $\Gamma_n$ the resonance width. 
The resonant states also satisfy the time-reversal property $u_{-n}(x)=u_{n}^*(x)$.
Since our approach is non-Hermitian, the resonant states  do not obey the usual rules of normalization and orthonormality \cite{gcaqc10}. Instead, the $u_n$'s fulfill the following condition,
\begin{equation}
\int_0^L u_n(x)u_m(x)\,dx +i \frac{u_n(0)u_m(0)+u_n(L)u_m(L)}{k_n+k_m}=\delta_{nm}.
\label{f5}
\end{equation}
For the particular case $n=m$, Eq.~(\ref{f5}) yields the corresponding  normalization for the resonant states,
\begin{equation}
\int_0^{L}u_{n}^2(x)\,dx + i \frac{u_{n}^2(0)+u_{n}^2(L)}{2k_n}=1,
\label{normalizationu}
\end{equation}
which also satisfy the closure relation,
\begin{equation}
\frac{1}{2}\,\sum_{-\infty}^{\infty}u_n(x)\,u_n(x')=\delta(x-x'),\,\,\,\, 0\le (x,x')^{\ddag} \le L,
\label{closure}
\end{equation}
where the notation $(x,x')^{\ddag}$ means that the expansion does not hold at $x=x'=0=L$.   
The resonant states  form a complete set \cite{gcaqc10} along the internal region of the potential, and any arbitrary function $\psi(x,k)$ can be expressed as a linear combination of these states using the fundamental relation $\psi(x,k)=2ikG^+(0,x;k)$. 

By using the  expansions (\ref{mittaglef}) and (\ref{Ggaston}) to evaluate the integrals in  Eq.~(\ref{integrales_complejas}) by a direct computation of the corresponding residues, we  obtain and  expression for $C(p,k)$ that can be substituted in Eq.~(\ref{psi_interanl1c}), leading us to,
\beqa
\widetilde{\Psi}(x,p)&=&\alpha\left[ \sum_{m=1}^{\infty}\frac{2 \alpha_m\kappa_{m}}{p \,(p-\kappa_{m})}G^+(0,L,\kappa_m) e^{-i \kappa_m L} e^{ipx}  \right. \nonumber \\
 &+& \left. \sum_{n=-\infty}^{\infty}\widetilde{\Phi}(0,k,k_n) \frac{e^{-i k_n L}u_n(0)u_n(x)}{p\,(p-k_n)} e^{i p x} \right].
\label{cuasifinal}
\eeqa     
We substitute in  Eq.~(\ref{cuasifinal})  the fundamental relationship between the transmission amplitude $t(k)$ and the Green function of the system \cite{gcaqc10}, namely
\begin{equation}
 t(k)=2 i k e^{-i k L}G^+(0,L;k),   
 \label{phiconG}
\end{equation}
and from the resulting expression we 
obtain $\Psi(x,t)$ by performing the inverse Laplace-transform  by means of the Bromwich integral, 
\begin{equation}
\displaystyle\Psi(x,t)=\frac{1}{2 \pi i}\int_{\gamma - i\infty}^{\gamma+i\infty}\widetilde{\Psi}(x;s)\,e^{st} ds.
\label{Bromwich_int}
\end{equation}
Equation~(\ref{Bromwich_int}) is evaluated by using the Moshinsky functions:
\beq
 M(x,q,t)=\frac{1}{2 \pi i} \int_{\gamma - i\infty}^{\gamma+i\infty}\frac{(i \alpha/2)}{\sqrt{\alpha is}}\left[ \frac{e^{\sqrt{\alpha i s }\, x}}{\sqrt{\alpha is }-q}\right]\, e^{s t} ds,
\label{Mcompeltota}
\eeq
which are also related to the complex error function  \cite{abramowitz}  $w(z)=e^{-z^2}{\rm erfc}(-i z)$, by
\begin{equation}
M(x,q,t)\equiv M(y_q)=\frac 1 2\,e^{imx^2/2 \hbar t}\, w(i y_q),
\label{moskyx}
\end{equation}
with,
\begin{equation}
y_q ={\rm e}^{-i\pi /4}\,\frac{\left(x- \hbar \, q \,t/m\right)}{\sqrt{2\hbar t/m}},
\label{yqx}
\end{equation}
where $q$ stands for $k$ or $k_{n}$.
We finally write the general time-dependent solution $\Psi(x,t)$ along the transmission region as: 
\beqa
\Psi(x,t)&=&- \Bigg[\sum_{m=1}^{\infty} \alpha_m t(\kappa_m) M(y_{\kappa_m})  \nonumber\\
&+&i\sum\limits_{n=-\infty }^{\infty
}t_n \widetilde{\Phi}(0,k,k_n)M(y_{k_n}) \Bigg],\,x\geq L.\nonumber \\
\label{eq:12}
\eeqa
In the above formula we have defined $t_n=e^{-i k_n L}u_n(0)u_n(L)$, and 
the $Q$-transform function $\widetilde{\Phi}(0,k,Q)$,
\begin{equation}
\widetilde{\Phi}(0,k,Q)=\Bigg[ i \int \phi(x,k)\, e^{-i Qx} dx\Bigg]_{x=0}.
\label{phitildosa}
\end{equation}
The coefficients $\alpha_m$ are the residues at the complex poles $Q=\kappa_m$ of  the $Q$-transform,  $\widetilde{\Phi}(0,k,Q)$. 
The solution given by Eq.~(\ref{eq:12}) holds for any finite range potential, and for initial states $\phi(x,k)$, as long as the $Q$-transform $\widetilde{\Phi}(0,k,Q)$ yields simple $Q$-poles.  
To verify  our result  Eq.~(\ref{eq:12}), let us consider as an example the case of an  absorbing plane wave shutter $\phi(x,k)=e^{ikx}$. The $Q$-transform is $\widetilde{\Phi}(0,k,Q)=(k-Q)^{-1}$ with one single pole at  $Q=\kappa_1=k$, with a residue $\alpha_1=-1$, that yields by direct substitution in Eq.~(\ref{eq:12})  the transmitted time-dependent wavefunction of Ref. \onlinecite{gcar97}. 

\subsection{Phase modulated wavepackets}

We use as an initial condition a cut-off phase modulated wavepacket 
given by,
\begin{equation}
\Psi(x,0)=\cases{2\,\sin (k\, x)\,\cos(\Delta k \,x), & for $x\leq 0;$\cr 0, & for $x> 0$,}
\label{eq1_bis_bis}
\end{equation}
constructed by a superposition of  quantum states with different momenta \textit{i.e.} 
$\Psi(x,0)= \psi_+(x,k_+,0)+\psi_-(x,k_-,0)$, with $\psi_{\pm}(x,k_{\pm},0)=(e^{i k_{\pm} x}-e^{-ik_{\pm} x})/2i$, where $k_{\pm}=k\pm\Delta k$, corresponding to different kinetic energies, $E_{\pm}=\hbar^2 k_{\pm}^2/2m$. 
The detailed analysis of the transient dynamics of these type of phase-modulated cut-off waves, to our knowledge, has not been performed before, although a similar condition has been applied in the context of time-dependent potentials \cite{Buttiker_1985}.

We apply our general result Eq.~(\ref{eq:12}) for a  modulated wavepacket $\phi(x,k)=2\,\sin (k\, x)\,\cos(\Delta k \,x)$, and  obtain,
\beq
\widetilde{\Phi}(0,k,Q)=i \Bigg[\frac{k_-}{(Q+k_-)(Q-k_-)}+\frac{k_+}{(Q+k_+)(Q-k_+)}\Bigg], \\
\label{qtransformodula}
\eeq
which has complex poles at $Q=\pm k_+$ (residues $\pm i/2$), and $Q=\pm k_-$ (residues $\pm i/2$). Thus, Eq.~(\ref{eq:12}) allows us to write,
\beq
\Psi(x,t)= \psi_+(x,k_+,t)+\psi_-(x,k_-,t); \,\,\,x\ge 0,
\label{psi_general_modula_suma}
\eeq
with
\beqa
\psi_{\pm}(x,k_{\pm},t)&=&(1/2i)\left[ t(k_{\pm})\,M(y_{k_{\pm}})- t(-k_{\pm})\,M(y_{-k_{\pm}}) \right.\nonumber \\
&+& \left. \sum\limits_{n=-\infty }^{\infty
}t_n \widetilde{\Phi}(0,k,k_n)M(y_{k_n}) \right].
\label{psi_general_modula}
\eeqa
In the next section  we shall apply our general result Eq.~(\ref{psi_general_modula_suma}) to explore the features of modulated wavepackets for a particular potential that involves a bound state.

\section{Dynamics of phase modulated wavepackets \label{results}}

Based on our result Eq.~(\ref{psi_general_modula_suma}), we shall explore the dynamics of the probability density of modulated cut-off wavepackets in quantum systems. 
As we shall show, the modulation of the initial state governs the  time-dependent features of the probability density in different regimes, and some of these features are independent of potential.
To begin our study, we consider the case of a simple potential that allows to explore the dynamics in systems with a single bound state, as is the case of the Dirac delta potential well, $V(x)=-\lambda \,\delta(x)$,  with $\lambda>0$. 
We obtain the corresponding solution from  Eqs.~(\ref{psi_general_modula_suma}) and (\ref{psi_general_modula}),  noting that only a single term in the resonance sum is required since the spectrum of the system is composed by a single bound state at $k_n=i \tilde{\lambda}$, with  energy $E_{b}=-\hbar^2\tilde{\lambda}^2/2m$, where  $\tilde{\lambda}=m\lambda/\hbar^2$. 
In the limit $L\rightarrow 0$, one can show with the help of Eq.~(\ref{normalizationu}) that $u_n^2(0)\rightarrow \tilde{\lambda}$, and thus $t_n\rightarrow \tilde{\lambda}$, leading us to the solution:
\beq
\Psi^{\delta}(x,t)= \psi_+^{\delta}(x,k_+,t)+\psi_-^{\delta}(x,k_-,t); \,\,\,x\ge 0,
\label{psi_delta_modula_2}
\eeq
with
\beqa
\psi_{\pm}^{\delta}(x,k_{\pm},t)&=& (1/2i)\left[ t(k_{\pm})\,M(y_{k_{\pm}})- t(-k_{\pm})\,M(y_{-k_{\pm}}) \right.\nonumber \\
&+& \left. \tilde{\lambda}\, \widetilde{\Phi}(0,k,i\tilde{\lambda})\,M(y_{i\tilde{\lambda}})\right],
\label{psi_delta_modula_2_bis}
\eeqa
where $t(k)=k/(k-i\tilde{\lambda})$
is the stationary transmission amplitude of the delta potential well. 
Also, the free modulated wavepacket solution is obtained directly from Eqs.~(\ref{psi_delta_modula_2}), and (\ref{psi_delta_modula_2_bis}) by letting $t(\pm k_{\pm})=1$, and $\tilde{\lambda}=0$, leading us to  
\beqa
\Psi_f(x,t)&=&(1/2i)[M(y_{k_+})+M(y_{k_-})\nonumber \\ 
&-& M(y_{-k_+})-M(y_{-k_-})], \,\,x\ge 0. \nonumber \\ 
\label{psi_delta_modula_free}
\eeqa
In what follows, we  use  Eq.~(\ref{psi_delta_modula_2}) for exploring  different regimes of probability density
$\rho(x,t)$ defined as,
\beq
\rho(x,t)= |\psi_+^{\delta}(x,k_+,t)+\psi_-^{\delta}(x,k_-,t)|^2; \,\,\,x\ge 0,
\label{psi_delta_modula_2_ah}
\eeq
to  investigate the effects of the phase modulation on the transient dynamics of the system.

\subsection{Quantum beats} \label{quantumbeats}

\begin{figure}[!tbp]
{\includegraphics[angle=0,width=3.4in]{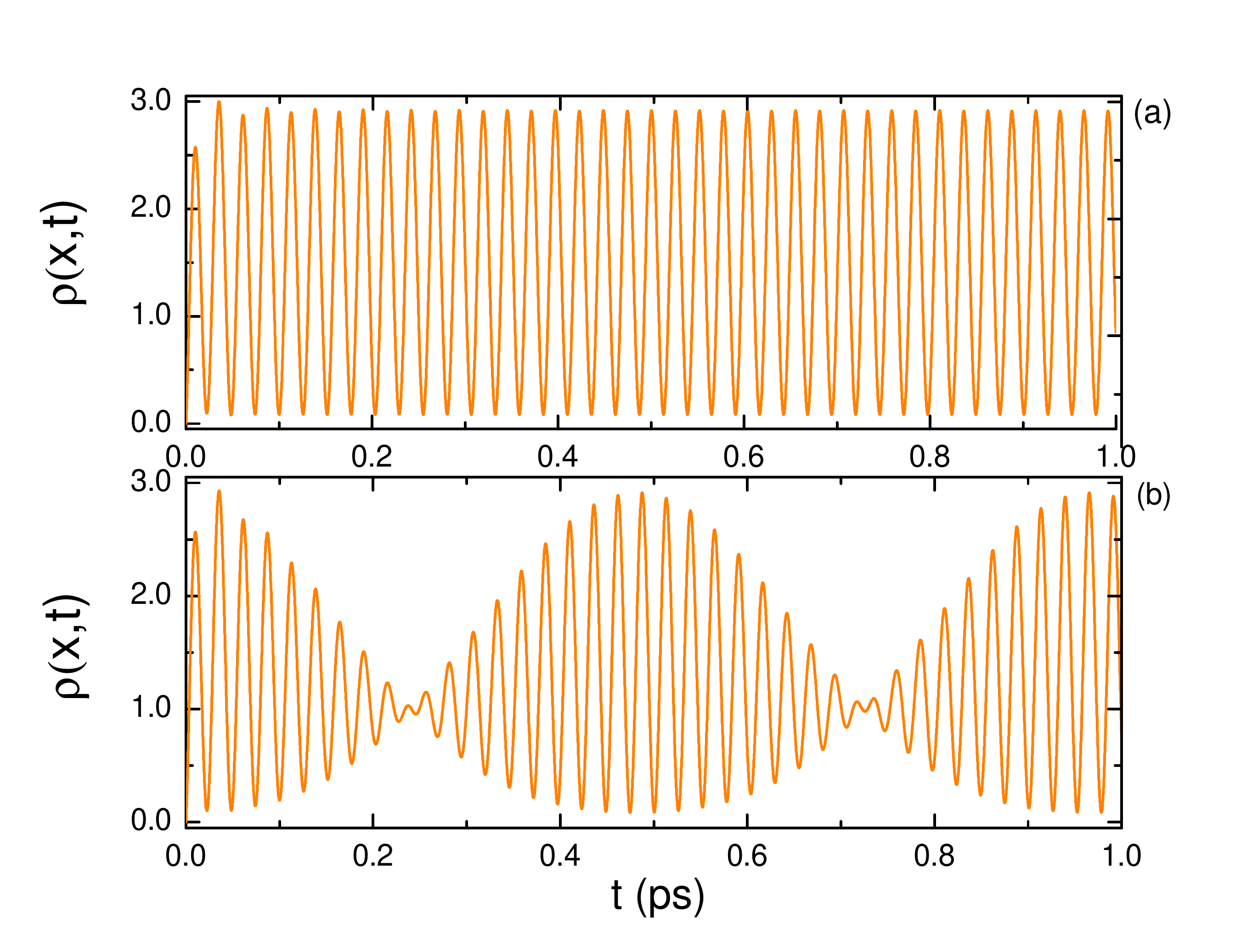}}
\caption{Probability density as function of time for a delta well potential at a fixed position $x=0$. In (a) we show that $\rho(0,t)$ [Eq.~(\ref{psi_delta_modula_2_ah})]  exhibits \textit{ persistent oscillations} for an unmodulated wavepacket  ($\Delta k=0$). In (b) we analyze the dynamics of the modulated probability density $\rho(0,t)$ for $\Delta k=0.001$ \AA$^{-1}$, where a \textit{beating effect} is observed. Here and in all the figures throughout the paper, we use the following parameters: delta intensity $\lambda=4.27$ eV \AA, and incidence energy $E=0.08$ eV. }
\label{qbshort}
\end{figure}

Let us now explore the time-dependent features of $\rho(x,t)$ [Eq.~(\ref{psi_delta_modula_2_ah})] at a fixed position $x=0$. In Fig. \ref{qbshort}(a) we plot $\rho(0,t)$ as function of  time, for an unmodulated wavepacket \textit{i.e.} $\Delta k=0$.
The probability density exhibits an oscillatory effect known as  {\it persistent oscillations}, reported in  Ref. \onlinecite{mlgc10} for the case of an absorbing plane wave shutter. 
The effect is a result of the interaction of the quasi-stationary state induced by the incident wave, and the bound state of system. 
We argue that the persistent oscillations also manifest themselves in a reflecting shutter setup, since the 
particular case with $\Delta k=0$ obtained from Eq.~(\ref{psi_delta_modula_2}), corresponds in fact (up to a phase factor $-i$) to a reflecting cut-off plane wave initial condition.
However, when we consider the case of a  modulated wavepacket ($\Delta k\ne0$), we show in Fig. \ref{qbshort}(b) that $\rho(0,t)$  exhibits high-frequency oscillations  in the time-domain, similar to a superposition effect of quantum waves with different frequencies. 
To identify the underlying superposition observed in Fig.~\ref{qbshort}(b), we plot the probability density for different values of $\Delta k$ over a larger time interval. 
\begin{figure}[!tbp]
{\includegraphics[angle=0,width=3.4in]{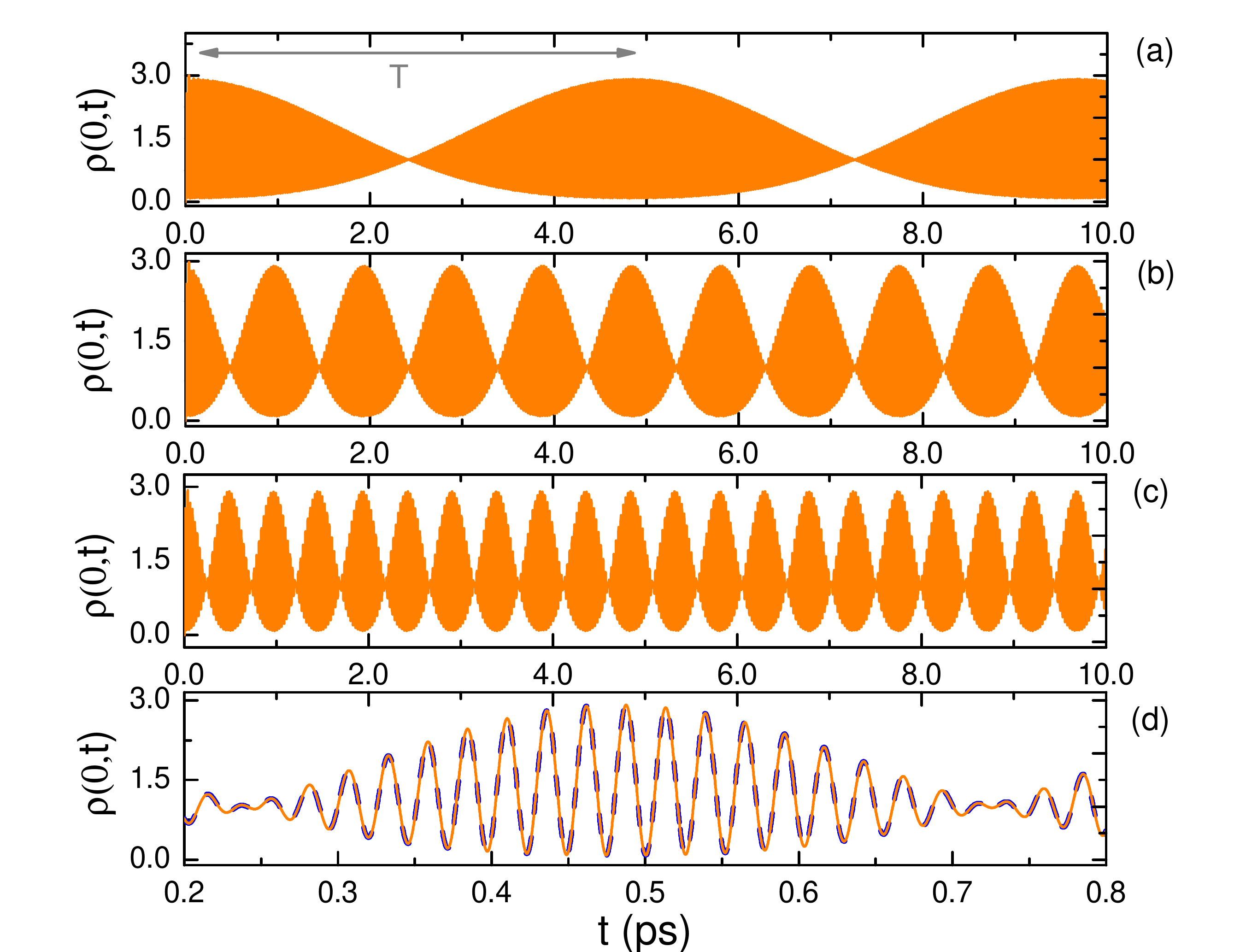}}
\caption{ \textit{Quantum beat} effect in the time-dependence of $\rho(0,t)$ [Eq.~(\ref{psi_delta_modula_2_ah})]  for a phase modulated wavepacket in a delta well potential, for (a) $\Delta k=0.0001$ \AA$^{-1}$, (b)  $\Delta k=0.0005$ \AA$^{-1}$, and (c) $\Delta k=0.001$ \AA$^{-1}$. 
The period of the beats is $T=4.9$ ps, as indicated in (a), corresponding to a beating frequency $\Omega= 1.3$ THz. (d) Comparison of the exact  $\rho(0,t)$ [Eq.~(\ref{psi_delta_modula_2_ah})] (orange solid line) of case (c), with $\rho_a(0,t)$ [Eq.~(\ref{chorizuda})] (blue dashed line), where a perfect agreement is observed.}
\label{qblong}
\end{figure} 
This is shown in Fig.~\ref{qblong}, where $\rho(0,t)$ exhibits \textit{quantum beats} with a well defined  frequency.
To determine the beating frequency of the oscillatory phenomenon, we derive an approximate solution for $\Psi^{\delta}(0,t)$  [Eq.~(\ref{psi_delta_modula_2})],  by using the properties of the Moshinsky functions.
We simplify our notation for the arguments of the Moshinsky function at $x=0$, by 
\begin{equation}
y(0,q,t) \equiv y^{\circ}_{q}=-{\rm e}^{-i\pi /4}\,\left( \hbar/2 m\right)^{1/2}q \,t^{1/2}.
\label{yq}
\end{equation}
By using the identity
$M(y^{\circ}_{q})={\rm e}^{-i \hbar q^2  t/2m}-M(y^{\circ}_{-q})$,
and keeping only the exponential contributions, we can approximate  $\Psi^{\delta}(0,t)$ by,  
\beqa
\Psi^{\delta}_{a}(0,t)&\simeq& \frac{1}{2 i} \left[t(k_+) e^{-i E_+ t/\hbar} +t(k_-) e^{-i E_- t/\hbar} \right . \nonumber \\
&+& \left. 2 \tilde{\lambda}\,\widetilde{\Phi}(0,k,i \tilde{\lambda})\, e^{-i E_b t/\hbar}\right].
\label{aproxenx0}
\eeqa
In Eq.~(\ref{aproxenx0}) we have neglected the contributions of Moshinsky functions of the form $M(y_{-q})$ since  these are fast decreasing functions of time.
The expression for the probability density $\rho_a(0,t)$ corresponding to Eq.~(\ref{aproxenx0}) is:
\beqa
\rho_a(0,t)&\simeq& \frac{1}{2}\Bigg\{ \frac{1}{2}\left[|t(k_+)|^2+|t(k_-)|^2+4 \tilde{\lambda}^2 |\widetilde{\Phi}(0,k,i \tilde{\lambda})|^2\right] \nonumber \\
&+&a_{1,r}\cos\Omega t+a_{1,i}\sin\Omega t \nonumber \\ &+&(a_{2,r}+a_{3,r})\cos\left(\bar{\Omega} t\right)\cos\left(\Omega t/2\right) \nonumber \\
&+&(a_{2,i}-a_{3,i})\sin\left(\bar{\Omega} t\right)\cos\left(\Omega t/2\right) \nonumber \\
&+&(a_{2,i}+a_{3,i})\cos\left(\bar{\Omega}  t\right)\sin\left(\Omega t/2\right) \nonumber \\
&+&(a_{3,r}-a_{2,r})\sin\left(\bar{\Omega} t\right)\sin\left(\Omega t/2\right) \Bigg\},
\label{chorizuda}
\eeqa
where the coefficients in Eq.~(\ref{chorizuda})  are calculated  as $a_{n,r}={\rm Re}[a_n]$, and $a_{n,i}={\rm Im}[a_n]$, with 
$a_1=t(k_+)t^*(k_-)$, $a_2=2t(k_+)\tilde{\lambda}\widetilde{\Phi}^*(0,k,i \tilde{\lambda})$, and  $a_3=2 t^*(k_-)\tilde{\lambda}\widetilde{\Phi}(0,k,i \tilde{\lambda})$.

We show from Eq.~(\ref{chorizuda}) that the dynamics of $\rho(0,t)$ in Figs.~\ref{qblong}(a)-(c) is governed by two types of frequencies, $\bar{\Omega}$ and $\Omega$, where 
$\bar{\Omega}=(\omega_++\omega_-)/2$, is an average frequency associated to  persistent oscillations, with 
$\omega_{\pm}= (E_{\pm}-E_b)/\hbar$.
The $\omega_{\pm}$ 
correspond to Rabi-type frequencies associated to the virtual states, $E_{\pm}$, induced by the source, and that of the bound state of the system, $E_b$.
The amplitude of the quantum wave is modulated by an envelope with a \textit{beating frequency}, $\Omega=\omega_+-\omega_-=(E_+-E_-)/\hbar$. Therefore,  the {\it quantum beats} in  Figs.~\ref{qblong}(a)-(c) arise due to the interaction between the {\it virtual two-level system} with energies $E_{+}$ and $E_{-}$, induced by the  modulation of the initial state.
In Fig.~\ref{qblong}(d) we compare the exact $\rho(0,t)$ [Eq.~(\ref{psi_delta_modula_2_ah})] with  $\rho_a(0,t)$  [Eq.~(\ref{chorizuda})], and a perfect agreement is observed. 
Note in  Fig.~\ref{qblong}(a) that the period $T=2\pi \,\Omega^{-1}$ indicated in the figure is of the order of picoseconds, and the  corresponding frequency of the \textit{quantum beats} lies in the terahertz range.  
Interestingly, the terahertz  frequency regime is currently accessible to experiments in the field  of  femtosecond pulse technology \cite{Garraway_1995}.

\subsection{Dynamics of a virtual two-level system}

\begin{figure}[!tbp]
{\includegraphics[angle=0,width=3.4in]{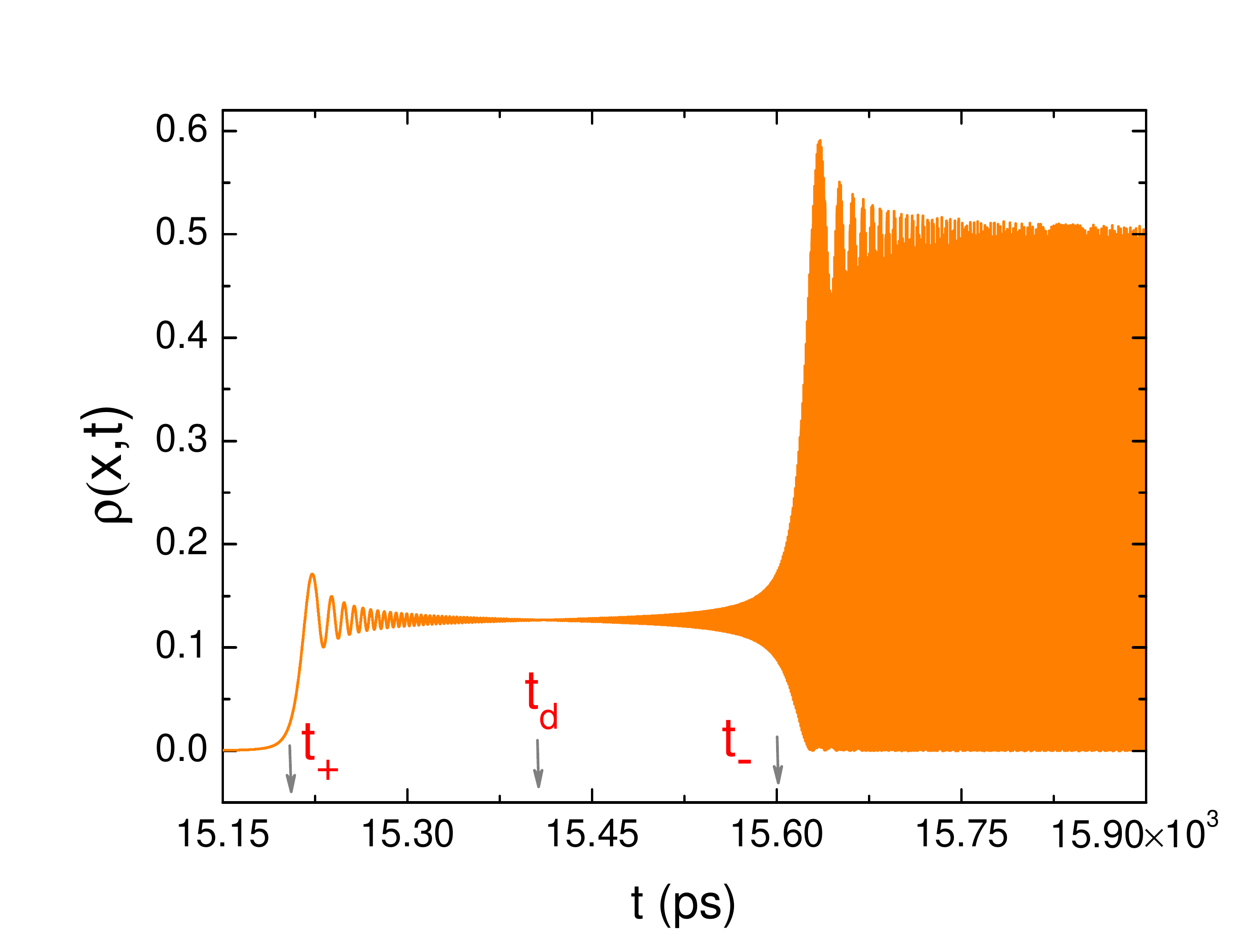}}
\caption{ Dynamics of  $\rho(x,t)$ [Eq.~(\ref{psi_delta_modula_2_ah})] at a fixed position $x=1\times10^8$ \AA~  with $\Delta k = 0.0005$ \AA$^{-1}$. We identify a regime governed by a {\it diffraction in time effect},  originated by the virtual level with energy $E_+$, which becomes important for times $t\simeq t_{+}$ onward. The time-diffraction pattern \cite{mm52} is observed over a time interval  $0\leq t \lesssim t_d$. At later times, for  $t>t_{-}$ onward, a highly-oscillatory interference pattern is observed.} 
\label{persistentes_grandes}
\end{figure}

\begin{figure}[!tbp]
{\includegraphics[angle=0,width=3.4in]{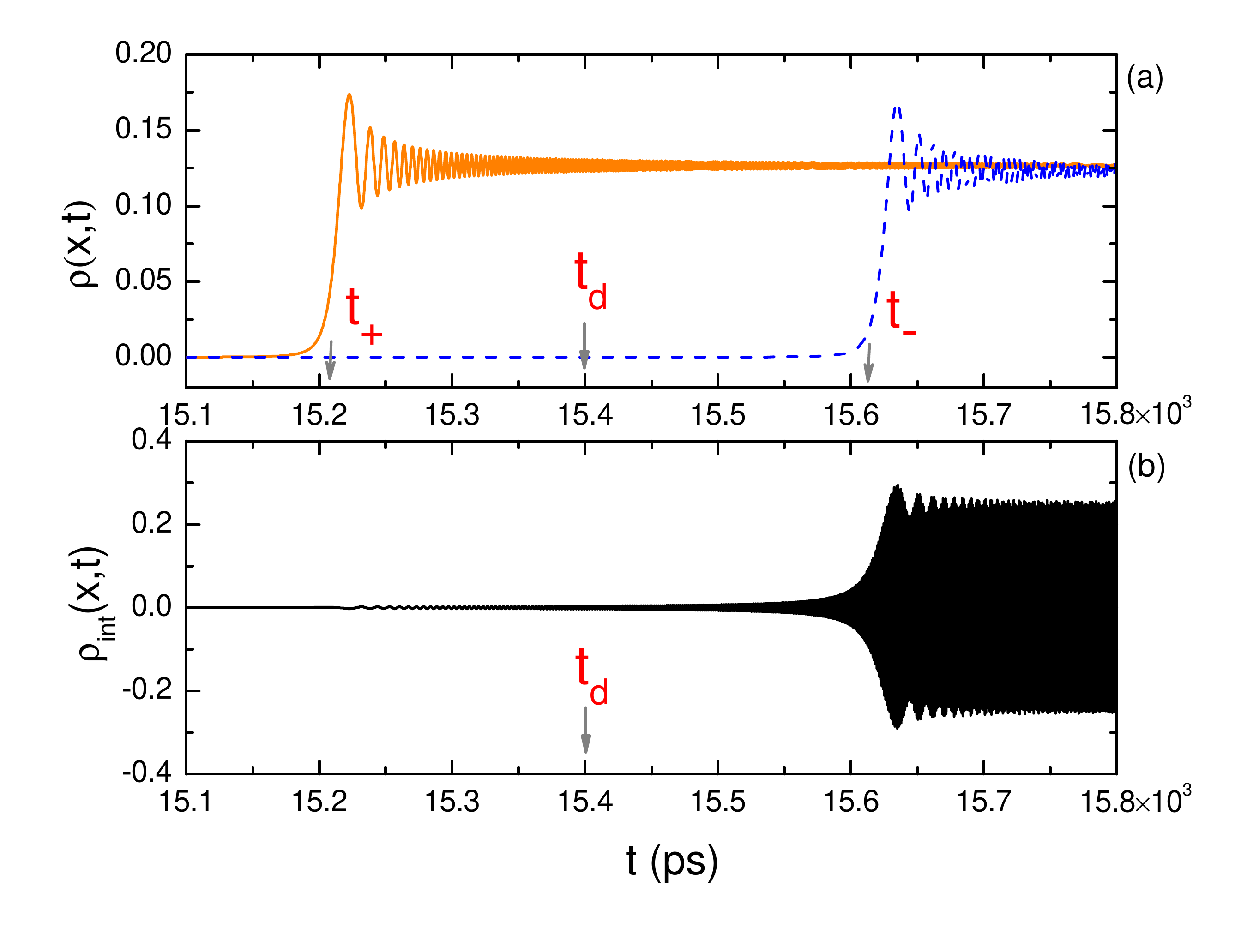}}
\caption{(a)  Time-dependence of the contributions $\rho_{\pm}$ to the exact $\rho(x,t)$ [Eq.~(\ref{psi_delta_modula_2_ah})] discussed in Fig.~\ref{persistentes_grandes}. Note that the arrival of the wavefronts of $\rho_{+}$ (orange solid line) and  $\rho_{-}$ (blue dashed line) occur at different times  $t_{+}$, and  $t_{-}$, respectively. We observe that $\rho_{\pm}$ exhibit a typical \textit{diffraction in time} phenomenon. (b) The interference $\rho_{int}$ of the contributions  $\rho_{\pm}$ becomes important for times $t>t_{-}$, setting the onset of the quantum interference between the virtual levels, $E_{\pm}$. The small interference between $\rho_{+}$ and $\rho_{-}$ over the time interval $0\leq t \lesssim t_d$, allows to observe the time-diffraction like pattern governed by $\rho_{+}$, as shown in Fig.~\ref{persistentes_grandes}.}
\label{persistentes_grandes_masmenos}
\end{figure}

We investigate the effect of the self-induced virtual levels, $E_{+}$ and $E_{-}$, on the dynamics of the probability density at distances far away from the potential.
These virtual levels arise due to the quantum superposition in our initial state, due to the phase modulation.  
This is illustrated in Fig.~\ref{persistentes_grandes}, where we show the time-evolution of $\rho(x,t)$ for large values of position, $x$.
The probability density features a sharp rise after the time of flight, $t=t_{+}\equiv x/v_+$, where $v_+=\hbar k_+/m$, giving rise to
a peculiar oscillatory-phenomenon known as \textit{diffraction in time} \cite{mm52}.
However, we note that from  $t\simeq t_{-}\equiv x/v_-$ onward, where $v_-=\hbar k_{-}/m$, the time-diffraction profile becomes a highly-oscillatory pattern.
The behavior observed in Fig.~\ref{persistentes_grandes} can be explained by analyzing the contributions to the exact solution $\Psi^{\delta}(x,t)$ [Eq.~(\ref{psi_delta_modula_2})] given by $\psi^{\delta}_{\pm}(x,t)$ [Eq.~(\ref{psi_delta_modula_2_bis})], and their corresponding interference. 
This is shown in Fig.~\ref{persistentes_grandes_masmenos}, where we present the contributions to the exact probability density by rewriting Eq.~(\ref{psi_delta_modula_2_ah}) as $\rho(x,t)=\rho_++\rho_-+\rho_{int}$, with $\rho_{\pm}=|\psi_{\pm}(x,k_{\pm},t)|^2$, and the interference term $\rho_{int}=2{\rm Re}[\psi_+(x,k_+,t)\psi_{-}^*(x,k_-,t)]$.
The time-dependence of $\rho(x,t)$ is characterized by the propagation of two wavefronts traveling with different speeds, $v_{+}$, and $v_{-}$. 
Since  $v_{+}>v_{-}$, it is expected that these traveling structures arrive at different times ($t_{+}<t_{-}$) at a fixed position, $x$.    
Therefore, in  Fig.~\ref{persistentes_grandes_masmenos}(a)  we argue that the dynamics of $\rho(x,t)$ for times $t\simeq t_{+}$, is characterized by a time-diffraction effect originated by the fast momentum components $k_+$, which is a manifestation of the virtual state with energy $E_+$.
At a later time, when $t\simeq t_{-}$, the slow momentum components $k_-$ begin to build the  virtual level of energy $E_-$.
From $t\gtrsim t_{-}$ onward, the interplay between the two self-induced virtual levels originates the highly-oscillatory profile observed in Fig.~\ref{persistentes_grandes}, due to the $\rho_{int}$ contribution shown in  Fig.~\ref{persistentes_grandes_masmenos}(b). 

\begin{figure}[!tbp]
{\includegraphics[angle=0,width=3.4in]{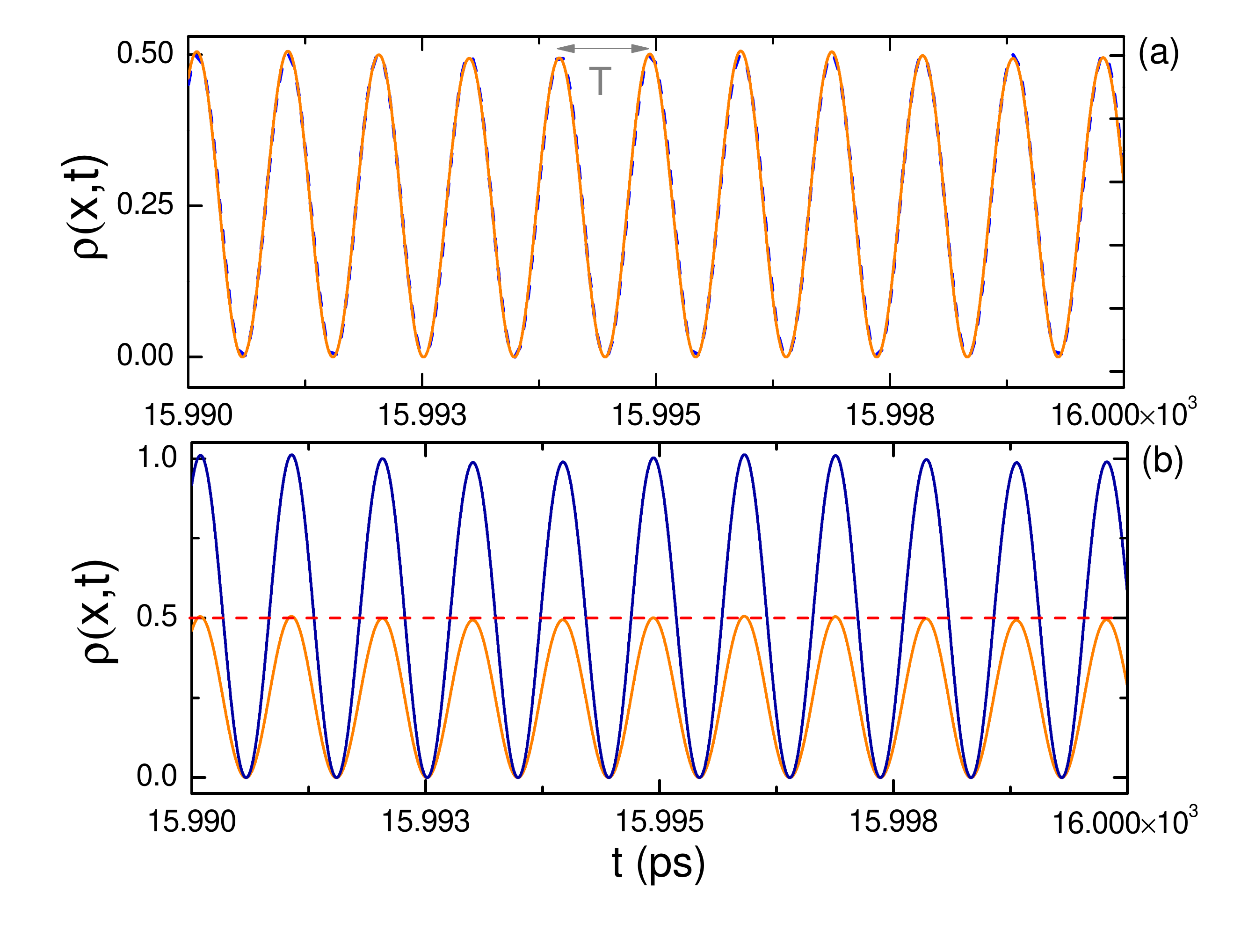}}
\caption{(a) Time-dependence of the exact $\rho(x,t)$ [Eq.~(\ref{psi_delta_modula_2_ah})] (orange solid line) for a delta potential well, and $\rho_{2l}(x,t)$ [Eq.~(\ref{chorizudita_bis})] (blue dashed line).  The \textit{virtual two-level system} dynamics  manifests itself as a Rabi oscillatory profile, with a frequency $\Omega \simeq 6.5$ THz and a period $T=0.97$ ps, indicated in the plot. (b) Comparison of the exact  $\rho(x,t)$ (orange solid line) discussed in (a), and the free case $\rho_{2l,f}(x,t)$ [Eq.~(\ref{rho_libriux})] (blue solid line). The case with $\Delta k=k$ (red dashed line) using  $\rho_{2l}(x,t)$ is also included, where the Rabi-mechanism is turned off.}
\label{qblargex}
\end{figure}

Next, we shall explore the effect of the virtual two-level system  in the long-time behavior of the probability density.
This is illustrated in Fig.~\ref{qblargex}(a), where we show that  the exact probability density [Eq.~(\ref{psi_delta_modula_2_ah})] exhibits periodic oscillations with a frequency $\Omega$. 
We shall show below that this transient feature is independent of the potential profile. 
We proceed by deriving an asymptotic solution from  Eq.~(\ref{psi_general_modula_suma}) with the help of the asymptotic properties of the $M$-functions. 
In the long-time limit ($t\rightarrow \infty$), and large values of the position, the argument of the $M$-functions is such that $|y_{q}|\gg 1$, leading to the following asymptotic behavior \cite{gcaqc10}:  
\beqa
M(y_q)&\simeq& \frac{e^{imx^2/2\hbar t}}{2\,\pi^{1/2}y_q}, \,\,-\pi/2<\mbox{arg}(y_q)<\pi/2; \label{moshyprop_a} \\ 
M(y_q)&\simeq& e^{imx^2/2\hbar t} e^{y_q^2}, \,\,\pi/2<\mbox{arg}(y_q)<3\pi/2.\label{moshyprop_b} 
\eeqa
By using the above expressions in Eq.~(\ref{psi_general_modula_suma}), we can see by inspection that the contributions involving the $M$-functions with $q=k_{n},-k_{\pm}$, are vanishingly small since $M(y_{q})\sim y_q^{-1}$  due to Eq.~(\ref{moshyprop_a}). 
Thus, only the $M$-functions with $q=k_{\pm}$, which are governed by Eq.~(\ref{moshyprop_b}), contribute to the dynamics, leading us to the result:  
\beq
\Psi_{2l}(x,t)\simeq \frac{1}{2 i} \left[t(k_{+}) e^{i(k_+ x -E_{+} t/\hbar)} +t(k_-) e^{i(k_{-} x -E_{-} t/\hbar)}\right]. \nonumber \\
\label{aproxenx0_bis}
\eeq
The corresponding probability density $\rho_{2l}(x,t)$ is given by,
\beqa
\rho_{2l}(x,t)&\simeq& \frac{1}{4}\bigg[ (|t(k_+)|+|t(k_-)|)^2 \nonumber \\ 
&-& 4|t(k_+)||t(k_-)|\sin^2\left(\Delta k x-\Omega t/2-\phi/2\right) \bigg], \nonumber \\
\label{chorizudita_bis}
\eeqa
with $\phi=\arctan(a_{1,i}/a_{1,r})$.
Note that $\rho_{2l}(x,t)$ exhibits an oscillatory profile typical of a two-level system  (Rabi-model), with a frequency  $\Omega$ (Rabi-frequency).
In Fig.~\ref{qblargex}(a) we compare  $\rho_{2l}(x,t)$ [Eq.~(\ref{chorizudita_bis})], with the exact  $\rho(x,t)$ for the delta well [Eq.~(\ref{psi_delta_modula_2_ah})], an a perfect agreement is observed.  
Furthermore, in Fig.~\ref{qblargex}(b) we show that the oscillatory phenomenon also manifests itself in the free probability density, computed from Eq.~(\ref{psi_delta_modula_free}), and given by 
\beqa
\rho_{2l,f}(x,t)\simeq 1-\sin^2(\Delta k\, x-\Omega t/2).  
\label{rho_libriux}
\eeqa
Our results in Fig.~\ref{qblargex}(b) also show that $\rho_{2l}(x,t)$ and  $\rho_{2l,f}(x,t)$ oscillate with the same frequency $\Omega$.
From a physical point of view, the Rabi-dynamics observed in the transient regime, is the result of the virtual \textit{self-induced two-level system}, with energies $E_+$ and $E_-$, where the frequency of oscillation $\Omega$ is proportional to the semi-classical propagation speed, $v_k$, and $\Delta k$ of the initial state, a feature that is independent of the potential. 
Only in the case where $\Delta k=k$, the oscillatory phenomenon disappears, as illustrated in Fig.~\ref{qblargex}(b). 
In this case, the effective cancellation of one of the channels occurs  \textit{i.e.} $E_{-}=0$, which hinders the Rabi-mechanism.
Alternatively, we can also argue that in the asymptotic regime our system behaves like a two-level system, by starting from Eq.~(\ref{aproxenx0_bis}) with the following change in notation,
\beq
\ket{\Psi(t)}= c_{+}e^{-iE_{+}t/\hbar}\ket{+} +c_{-} e^{-iE_{-} t/\hbar}\ket{-}, 
\label{psiket_a}
\eeq
where $\{\ket{+}$, $\ket{-}\}$ is an orthogonal base with eigenstates  $\{E_{+}$, $E_{-}\}$, and the coefficients $c_{\pm}$ take into account the $t(k_{\pm})$, and other factors.  
By eliminating the phase factor $e^{-iE_{-}t/\hbar}$ from Eq.~(\ref{psiket_a}), we can obtain 
%
the probability of finding the state $\ket{\Psi(t)}$ at an earlier time $t=t'$ which is given by  $P_{t,t'}=|\braket{\Psi(t')|\Psi(t)}|^2$, where the system was initially at one of the virtual states ($E_+$ or  $E_-$), leading us to
\beqa
P_{t,t'}=(|c_+|^2+|c_-|^2)^2-4|c_+|^2|c_-|^2\sin^2\left[\Omega (t-t')/2\right], \nonumber \\
\label{psiket_c}
\eeqa
which is the well known Rabi-formula.  
These persistent asymptotic oscillations are a feature of the initial modulated state, and are present for any finite range potential.

\subsection{Delay-time for modulated  wavepackets \label{timedelaysec}}

\begin{figure}[!tbp]
{\includegraphics[angle=0,width=3.4in]{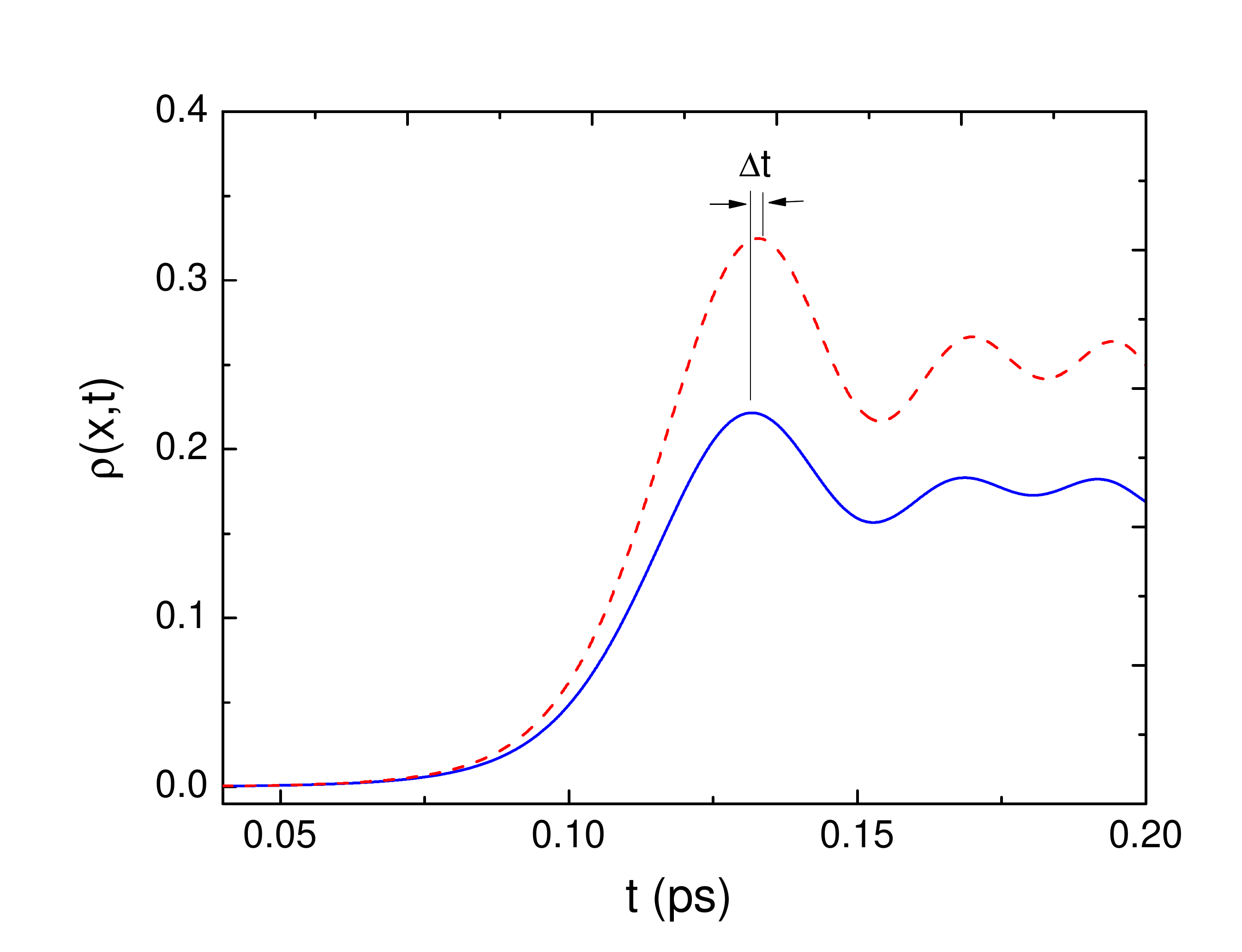}}
\caption{Measurement  of the delay-time $\Delta t$ by considering plots of $\rho(x,t)$ for a delta well  [Eq.~(\ref{psi_delta_modula_2_ah})]  (blue solid line), and the free-case $\rho_f(x,t)$  (red dashed line), at fixed position $x=1000.0$ \AA, and $\Delta k=0.02$  \AA$^{-1}$. The corresponding delay-time is $\Delta t=-1.055\times10^{-3}$ ps.}
\label{delaytime}
\end{figure}

\begin{figure}[!tbp]
{\includegraphics[angle=0,width=3.4in]{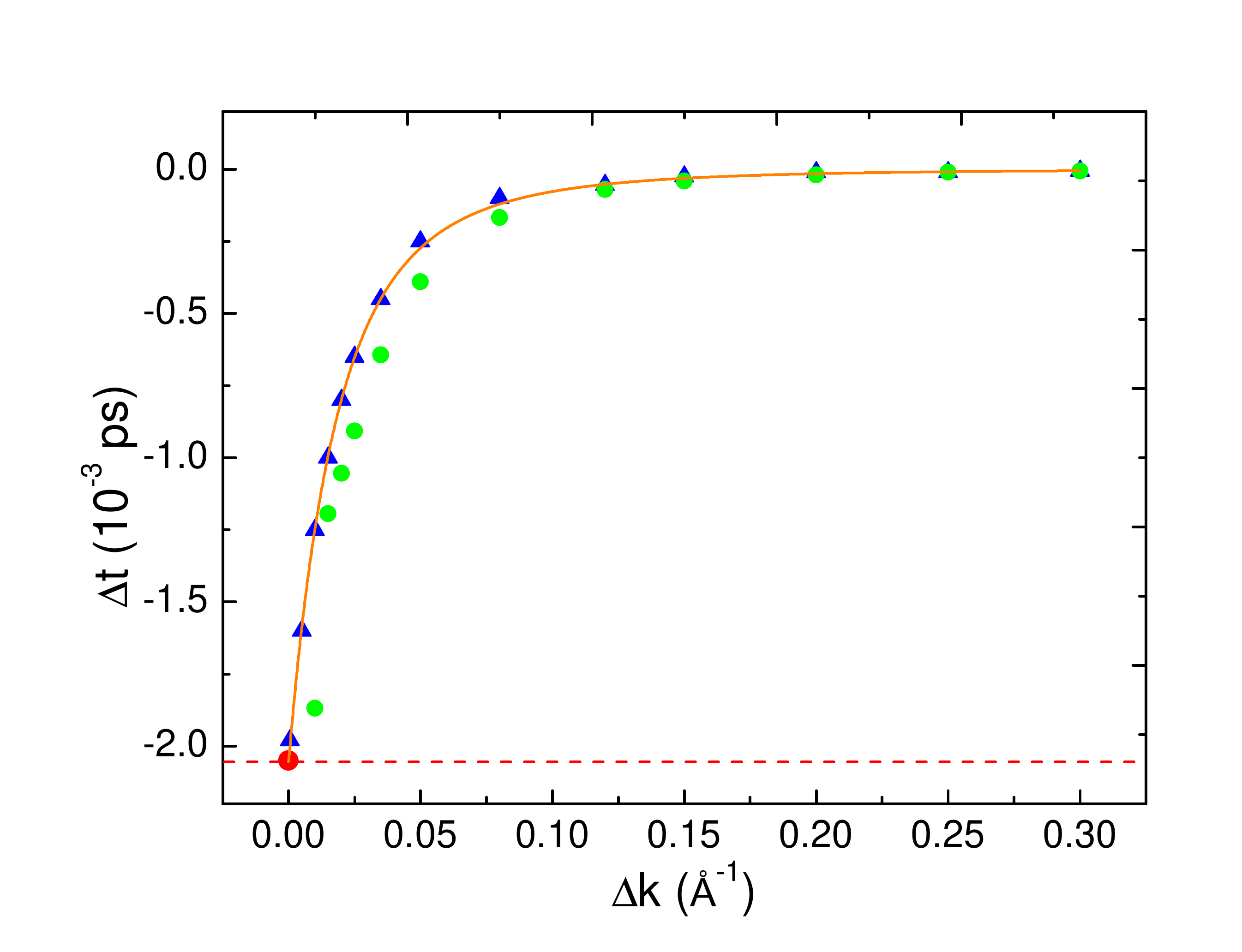}}
\caption{Delay-time $\Delta t$ as function of $\Delta k$ for the modulated wavepacket for a delta potential well,  calculated using Eq.~(\ref{delay}), measured at fixed positions $x= 1.0 \times 10^8$  \AA ~(blue solid triangles), and $x=1000.0$ \AA ~(green solid circles).  A plot of the analytical expression for the delay-time Eq.~(\ref{delya_anal})  (orange solid line) is also included for comparison, and an excellent agreement is observed. The red solid circle corresponds to delay-time for the case of $\Delta k=k$. The latter corresponds to a plane wave. The red dashed line corresponds to the phase-time Eq.~(\ref{tphi}), which yields $t_{\phi} = -2.05  \times 10^{-3}$ ps.}
\label{delayvsdeltak}
\end{figure}

The delay-time \cite{DECARVALHO200283,mohsen2013quantum,Muga:2023382} as a 
transient effect has been observed in quantum structures, such as  resonant double barrier systems \cite{gcar97}, and delta potentials \cite{19306687,gcah03,19156601,mlgc10,PhysRevA.86.062118}, and depending of the system, this time scale may exhibit advance or delay-time. 
For example, in the case of repulsive delta potentials \cite{gcah03},  a delay-time is observed, while in the case of attractive potentials \cite{mlgc10}, a time-advance is obtained.
It is the purpose of this section to study the effect of the \textit{virtual states}, $E_+$, and $E_-$, on such a transient time-scale.
We calculate the \textit{dynamical delay-time} $\Delta t$ following the prescription of previous work on this subject \cite{gcah03}, which involves the measurement of the maxima of the main wavefront of the probability density in the time-domain. These maximum values of  $\rho(x,t)$ for the delta well  [Eq.~(\ref{psi_delta_modula_2_ah})], and the free-case $\rho_f(x,t)=|\Psi_f(x,t)|^2$ using Eq.~(\ref{psi_delta_modula_free}), occur at $t_{\delta}$ and $t_{f}$, respectively, where the delay-time, $\Delta t$, corresponds to the difference,
\begin{equation}
\Delta t\equiv t_{\delta}-t_{f},
\label{delay}
\end{equation}
as illustrated in Fig.~\ref{delaytime}.
Note also in this figure, that at a fixed position, $x$, the maximum of the probability density for the delta well case appears before the free-propagation peak, leading to a time-advance. 
In both cases,  the dynamics is governed by the virtual level, $E_+$.
In Fig.~\ref{delayvsdeltak} we present the measurements of $\Delta t$ as function of $\Delta k$, for two different values of the position, $x$, far from the potential.
We show in Fig.~\ref{delayvsdeltak} that for small values of  $\Delta k$, the delay-time is very close to phase time. 
In contrast, the delay-time tends to zero for large values of $\Delta k$
because the initial state is very energetic, and it is not affected by the potential.

We derive an exact analytical expression for $\Delta t$ by 
following the procedure of Ref. \onlinecite{PhysRevA.86.062118}, which yields the result:
\beq
\Delta t \simeq \frac{(-\lambda/2)}{v_{+}\,\left(E_{+}-E_{b} \right)}.
\label{delya_anal}
\eeq
%
%
The time-scale Eq.~(\ref{delya_anal}) is explicitly governed by the energies $E_+$ and $E_b$, and is also included in Fig.~\ref{delayvsdeltak}, where an excellent agreement is observed.
We also compute the phase-time $t_{\phi}$ of the system by using the definition:
\begin{equation}
t_{\phi}=\hbar\,\frac{d\phi_t}{dE}=\hbar \,\textrm{Im} \left[ \frac{1}{t(k)} \frac{dt(k)}{dk} \frac{dk}{dE}  \right].
\label{delay_definition_a}
\end{equation}
By substituting in Eq.~(\ref{delay_definition_a}) the transmission amplitude, $t(k)$, for the delta potential, we obtain the explicit phase-time,
\beq
t_{\phi}= \frac{(-\lambda/2)}{v_k\,\left(E-E_b \right)}.
\label{tphi}
\eeq
Note that in the limit $\Delta k\rightarrow 0$, $E_+\rightarrow E$, and $v_+\rightarrow v_k$, thus $\Delta t \rightarrow t_{\phi}$, that is, the phase-time is recovered for a quasi-monochromatic initial state.

\section{Conclusions}
\label{conclusions}
We explore the transient dynamics of phase modulated cut-off wavepackets by deriving an exact analytical solution to Schr\"odinger's equation for  finite range potentials, within a quantum shutter approach involving a general initial quantum state. 
The  modulated  cut-off quantum wave allows us to explore matter-wave transient phenomena such as time-diffraction, wave superposition, quantum beats, Rabi-oscillations, and delay-time. 
In particular, it is demonstrated that the time-dependent features of the probability density are governed by a virtual \textit{self-induced two-level system}, with energies $E_+$ and $E_-$, due to the modulation of the initial state. 
We found that in the long-time regime  the probability density exhibits Rabi-dynamics which is independent of the potential profile, with a frequency $\Omega=(E_+-E_-)/\hbar$.  Interestingly, since $\Omega=2 v_k \Delta k$, where $v_k$ is the velocity of a plane wave  with momentum $k$, the Rabi-frequency can be tuned by controlling the incidence energy $E$ and $\Delta k$ of the modulated wave.
The Rabi-mechanism can be effectively turned off by setting $\Delta k=k$, which corresponds to the closing of the channel $E_-$, or by letting $\Delta k=0$ (unmodulated case).
For the particular case of a  system with a bound state, $E_b$, we found that the virtual two-level system gives rise (at $x=0$) to a series of \textit{quantum beats} in the probability density, with a beating frequency $\Omega$. 
These quantum beats in the probability density modulate a highly-oscillatory pattern of average frequency $\bar{\Omega}$ associated to persistent oscillations, resulting from an interplay between the virtual states, and the bound state of the system. 
We also demonstrate the existence of a transient regime governed by $E_+$, where the probability density exhibits a \textit{diffraction in time} effect, which let us identify a well defined traveling wavefront. This  feature allows to measure the delay-time $\Delta t$ of the system at far distances from the potential. We also derive a simple analytical formula for describing this time scale, which depends on the energy difference between $E_+$ and $E_b$.
It is also shown that only for the case of plane waves, $\Delta t$ agrees with the phase-time.
The latter is in agreement with the results obtained for the delay time in Refs. \onlinecite{gcah03} and \onlinecite{mlgc10}, involving the time-evolution of cut-off plane waves.

We stress that both the \textit{quantum beat} effect, as well as the Rabi oscillations,  occur in a frequency range of the order of terahertz, which is currently accessible to experiments involving femtosecond spectroscopy  \cite{Garraway_1995}. 
These matter-wave phenomena associated to electrons can be explored as discussed in Ref. \onlinecite{decamps16}, by using interferometric techniques based on electron holographic microscopy \cite{Hasselbach_2009}.

\section{Acknowledgements}
\label{acknowledgements}

The authors acknowledge support from UABC under Grant PFCE 2018.

\end{document}